\input epsf
\documentclass[nohyper,notoc]{article} 
\usepackage{epsfig}

\def\beq{\begin{equation}}
\def\eeq{\end{equation}}
\def\bea{\begin{eqnarray}}
\def\eea{\end{eqnarray}}
\def\bq{\begin{quote}}
\def\eq{\end{quote}}

\def\gappeq{\mathrel{\rlap {\raise.5ex\hbox{$>$}}
{\lower.5ex\hbox{$\sim$}}}}
\def\lappeq{\mathrel{\rlap{\raise.5ex\hbox{$<$}}
{\lower.5ex\hbox{$\sim$}}}}

\def\mec{$\mu - e$ conversion}

\def\Aslash{A \!\! \! \!   /~}
\def\Eslash{E \!\! \! \!   /~}

\def\sslash{s \! \! \!   /~}

\def\kslash{k \! \! \!   /~}
\def\smu{s_\mu \! \!\!  \!  \!   /~}
\def\se{s_e \! \!\!  \!  \!   /~}

\begin{document}

\renewcommand{\thefootnote}{\fnsymbol{footnote}}

\begin{center}
{\Large {\bf CP violating phases  in  $\mu-e$ conversion
}}
\vskip 25pt
{\bf 
 Sacha Davidson 
\footnote{E-mail address:
s.davidson@ipnl.in2p3.fr}
 }
\vskip 10pt  
{\it IPNL, Universit\'e Lyon 1, CNRS/IN2P3, 69622 Villeurbanne cedex, France }\\
\vskip 20pt
{\bf Abstract}
\end{center}

\begin{quotation}
  {\noindent\small 

Experiments are planned to improve the
sensitivity of \mec ~ from  the current $\sim 10^{-12}$ to
$\sim 10^{-16} - 10^{-18}$.  If the muon (bound
to the nucleus) could be polarised, a spin asymmetry
of the final state electron is  sensitive to
CP violating phases on lepton flavour violating
operators. This is similar to extracting
phases from  asymmetries in the final state  
spin  and phase space distributions
of   $\mu \to 3e$ and $\mu \to e \gamma$.

\vskip 10pt
\noindent
}

\end{quotation}

\vskip 20pt  

\setcounter{footnote}{0}
\renewcommand{\thefootnote}{\arabic{footnote}}

\section{Introduction}
\label{intro}

CP is a discrete  transformation,
turning particules into anti-particles. 
In the Lagangian,   it can
be implemented by taking  the complex conjugate
of all the coupling constants.
 So CP Violation(CPV)  can arise 
when the coupling constants have unremoveable phases. 
 In the  quark sector of the SM,  there is
one  such unremoveable phase, and it is observed to
be of order one.   However, if there is
new physics at the weak scale,  the non-observation
of  electric dipole moments implies 
that  combinations of the new physics
phases must be small.  The origin of CP Violation
thus remains an enigma---are all coupling
constants equiped with ${\cal O}(1)$ phases, or
is CPV a particularity of the quark sector?

 A source of CPV, beyond  the phase of
the quark mixing matrix, seems required to 
generate the excess of matter over anti-matter
observed in the Universe.  Some new
physics is also required to generate neutrino
masses, so the seesaw mechanism \cite{seesaw} is a 
popular extension of the SM because it
naturally generates the observed small
neutrino masses and can  produce the
baryon asymmetry  via  thermal leptogenesis \cite{FY,review}.
For this to occur,  CP Violating  phases 
are required in the lepton sector.

One of the major attractions of proposed
third generation neutrino beam facilities \cite{iss}
(neutrino factory, $\beta$-beam or superbeam),
is their sensitivity to the phase $\delta$ of
the lepton mixing matrix. Unfortunately
these are expensive machines.  So  it is interesting
to enquire if leptonic CP Violation can be found somewhere else.

This letter considers  the sensitivity of 
 $\mu-e$ conversion  \cite{KunoOkada} to the phases
of dipole interactions (see eqn (\ref{eqn2})). This
would be ambiguous evidence for CP Violation in the
lepton sector,  because   Lepton
Flavour Violating (LFV)  processes  like
$\mu-e$ conversion require new
physics at the electroweak scale,
which  may not be the same new physics
as generates neutrino masses. For instance,
in the supersymmetric seesaw,  the
CPV phases appearing in $\mu-e$ conversion
could arise from the neutrino Yukawa couplings
via renormalisation group running, or could be
intrinsic to the  soft supersymmetry-breaking
 parameters. Nonetheless,
CPV is elusive, so any observation is interesting.

The asymmetries we are interested in
arise from  ``triple products'' of
spin and/or momentum three vectors
\cite{tp}, and  are sometimes refered to as
T-odd asymmetries. Technically, they arise 
when the  $i$ from the Dirac trace  $\gamma^{\alpha} \gamma^{\beta } 
\gamma^{\sigma } \gamma^{\delta} \gamma^{5} = 
-4 i \varepsilon^{\alpha \beta \sigma \delta}$,
multiplies  a  CPV phase
from the coupling constants (the case
we are interested in), or   the Imaginary part of an amplitude 
($eg$ induced by on-shell intermediate
particles in a loop).  Such   a ``triple product''
asymmetry could manifest itself, 
for instance,   as   a forward-backward
asymmetry in  decays of a particle. 
Notice that it does not require the simultaneous
presence of an phase in the amplitude and
the coupling constants, as 
is required  to obtain  a difference 
between the 
integrated (over final
state momentum and spin) 
decay rates of the particle and antiparticle
\footnote{ $S-$matrix unitarity relates
$\langle (\vec{p}_1,\vec{s}_1)... (\vec{p}_n,\vec{s}_n)
| (\vec{k}_1,\vec{s}'_1)... (\vec{k}_m,\vec{s}'_m) \rangle $
to 
$\langle (\vec{k}_1,\vec{s}'_1)... (\vec{k}_m,\vec{s}'_m) 
| (\vec{p}_1,\vec{s}_1)... (\vec{p}_n,\vec{s}_n) \rangle $. 
$T$  transforms 
$\langle (\vec{p}_1,\vec{s}_1)... (\vec{p}_n,\vec{s}_n)
| (\vec{k}_1,\vec{s}'_1)... (\vec{k}_m,\vec{s}'_m) \rangle $ $
\to $ 
$\langle (-\vec{k}_1,-\vec{s}'_1)... (-\vec{k}_m,-\vec{s}'_m) 
| (-\vec{p}_1,-\vec{s}_1)... (-\vec{p}_n,-\vec{s}_n) \rangle $.}.

Asymmetries which are sensitive to
CPV phases, that can arise  in $\mu \to 3e$,
 are reviewed in \cite{KunoOkada}
and discussed in  \cite{CERNreport}.  
Farzan \cite{Farzan} showed recently
that  CPV phases could be
measured in  $\mu \to e \gamma$, 
if the  $\gamma$  and $e$ polarisation
could be measured.
Here
we focus on $\mu e$ conversion.
 The current bounds on Titanium ($Z = 22$) and
gold ($Z = 79$)  from Sindrum2 at PSI
\cite{Bertl:2006up}  are
\beq
\frac
{\Gamma( \mu {\rm Ti} \rightarrow e {\rm Ti})}
{\Gamma( \mu {\rm Ti}  \rightarrow capture)}  < 4.3 \times 10^{-12} 
~~~~,~~~~~~~~
\frac
{\Gamma( \mu {\rm Au} \rightarrow e {\rm Au})}
{\Gamma( \mu {\rm Au}  \rightarrow capture)}  < 7 \times 10^{-13} ~~.
\eeq
Recall that the branching ratio scales $\propto Z$\cite{KunoOkada}.
There are experiments under discussion 
($\mu2e$ at FermiLab\cite{Carey:2007zz}, 
PRISM/PRIME \cite{PP} at J-PARC) aiming for
sensitivities of $10^{-16} - 10^{-18}$. 
This letter  assumes 
that  the muon can be 
 polarised, which is not automatic;  the possibility of
polarising the muon is discussed in
\cite{Kuno:1987dp}.

  \section{ \mec}

Consider the effective dipole interaction, between a
$\mu$, an $ e$  and a photon, described by:
\beq
-\frac{4 G_F m_\mu}{\sqrt{2}}
\overline{\mu} \sigma_{\mu \nu} (A_L P_R + A_R P_L) e F^{\mu \nu} + h.c
\label{eqn2}
\eeq
where $\sigma_{\mu \nu} = \frac{i}{2} [\gamma_\mu, \gamma_\nu]$,
and the potentially complex  dimensionless  coefficients $A_L$ and $A_R$ 
arise by integrating out   new physics
in the vertex correction \cite{dipolelh,dipolesusy,Okada2000,dipoleunpart}.
In addition of eqn (\ref{eqn2}), New Physics can induce
monopole  $(\propto q^2 \bar{\mu} \Aslash e)$ and 
four-fermion interactions, which are neglected here.  
In some (for instance, supersymmetric) models, the
dipole interactions are the most significant. 
The interactions of eqn (\ref{eqn2})  allow the
decay of a
$\mu$ of momentum and spin $p, s_\mu$   to 
an $ e$ of momentum and spin $k, s_e$, and a  $ \gamma$
 of momentum $q$. 
The total branching ratio, integrated over phase space and
summed on spins, is insensitive to
the phases on $A_L$ and $A_R$: 
\beq
BR (\mu \to e \gamma) = 384 \pi^2 (|A_L|^2 + |A_R|^2)
\eeq

To find triple products which are  sensitive to CPV phases,
consider instead the differential rate, or
better the unintegrated  $|$matrix element$|^2$,
not summed over spins. It must be Lorentz invariant.
If  one can construct  a Lorentz invariant CP-odd
combination of spins and momenta, then one could
enquire if it multiplies the  potentially 
CP-odd combination $A_L A_R^*$.  If yes, then
there would be a CP  asymmetry in the differential
rate.  CP odd, Lorentz invariant combinations
of 4-vectors could be
\beq
\epsilon_{\alpha \beta \rho \sigma} 
p_\mu^{\alpha}k^{ \beta} s_\mu^ \rho s_e^\sigma ~~~,~~
\epsilon_{\alpha \beta \rho \sigma} 
p_\mu^{\alpha}q^{ \beta} s_\mu^ \rho s_e^\sigma 
\eeq
These can appear in the Dirac traces,  in
the presense of a $\gamma_5$. 
In the $|$matrix element$|^2$, appear
always two powers of the photon four-momentum  $q$, so the first possibility
is multiplied by $q^2 = 0$. The second
possibility does not appear.   It was recently shown by Farzan
\cite{Farzan},
that one must also include the spin of the photon,
if one wishes to find a CP asymmetry
in the $\mu \to e \gamma$ decay.  Alternatively,
one can study  the decay $\mu \to 3e$,
which has $q^2 \neq 0$, where there is an asymmetry
sensitive to the relative phase between  $A_L$ and $A_R$\cite{Farzan}
\footnote{Other  triple product asymmetries 
in $\mu \to 3e$ are sensitive to other New
Physics phases\cite{Okada2000}} .
Also in $\mu -e$ conversion, $q^2 \simeq m_\mu^2$ \cite{czar}.

In \mec ~ experiments, an incident $\mu^-$ is captured
by a nucleus, then cascades down to the $1s$ state. 
The muon usually turns into a neutrino, by
muon capture $\beta$ decay, or it could convert
to an electron in the presence of New Physics. 
A detailed discussion of how to compute \mec ~rates,
from effective interaction including  eqn (\ref{eqn2}),
can be found in \cite{czar,Kitano}. 
In the approximation that (\ref{eqn2}) is
the only interaction permitting $\mu -e$
flavour change,  the transition takes place in the electric
field of the nucleus, with a   matrix element
 \cite{Kitano}
\beq
{\cal M} ={\cal M}_L + {\cal M}_R =  -
\frac{4G_F m_\mu}{\sqrt{2}} \int d^3 x
e^{-i \vec{k} \cdot \vec{x}}
\left( A^*_L \overline u(k) 2 \sigma^{0i} E_i
P_L{\psi}_{1s}^{(\mu)}
+ 
 A^*_R \overline u(k) 2 \sigma^{0i} E_i
P_R{\psi}_{1s}^{(\mu)}
\right)
\eeq
where $\vec{E}$ is the electric field of
the nucleus, ${\psi}_{1s}^{(\mu)}$ is the 
wavefunction of the muon in the $1s$ state,
and the electron has been approximated  (as in \cite{FW})
as a free plane wave  of momentum $\vec{k}$. 
 This approximation
is less good for heavy nuclei \cite{czar,Kitano}.

Define  
the average  over the muon wavefunction:
$$ 
 \int d^3 x
{\psi}_{1s}^{(\mu)} E^i  \equiv \langle E^i \rangle ~~~, 
$$
then  using $2\langle E \rangle_i \sigma^{0i} =i \gamma^0 \langle \Eslash \rangle$,
 defining $\langle \Eslash \rangle^* = (\langle E \rangle^i)^*\gamma_i$
and neglecting  $m_e$, one obtains
\bea
\frac{|{\cal M}|^2}{8G_F^2 m_\mu^2 }& =& {\cal M}_L  {\cal M}_L^*+
{\cal M}_L^*  {\cal M}_R+
{\cal M}_L {\cal M}_R^* + {\cal M}_R {\cal M}_R^*\nonumber \\
 &
= & | A_L|^2 {\Big \{ } \kslash ~ \gamma^0 \langle \Eslash \rangle  P_L 
\gamma^0 P_R \gamma^0 \langle \Eslash \rangle^* {\Big \} } +
 | A_R|^2 {\Big \{ } \kslash ~   \gamma^0 \langle \Eslash \rangle  P_R 
\gamma^0 P_L \gamma^0 \langle \Eslash \rangle^* {\Big \} }
\nonumber \\
&&+ 
A_LA_R^* {\Big \{ }\kslash ~ 
 \gamma^0 \langle \Eslash \rangle  P_R 
\gamma^0 P_R \gamma^0 \langle \Eslash \rangle^* {\Big \} }
+ A_L^*A_R
 {\Big \{ }\kslash ~ 
 \gamma^0 \langle \Eslash \rangle P_L
\gamma^0 P_L \gamma^0 \langle \Eslash \rangle^* {\Big \} }
\eea
 where inside curly brackets
one should take a Dirac trace.
The $A_L A_R^*$ cross terms drop out because
of the chirality
projection operators, and with  $k_0 = m_\mu$, 
this gives 
\beq
\Gamma =  64 \pi G_F^2 m^3_\mu  |\langle E \rangle|^2    ( |A_R|^2+ |A_L|^2) 
\label{total}
\eeq


The conversion rate
for polarised muons and electrons, can be obtained
using   the  polarisation projection
operators:
\beq
\frac{1}{2} \left( I +   \gamma_5 \sslash \right)
\eeq
(where $s_\alpha$ is the spin  of particle $\alpha$)
to the wavefunctions. The $A_L A_R^*$  part of
$|{\cal M}|^2$ is
\bea
{\cal M}_L {\cal M}_R^* + h.c. & =
& - {2G_F^2 m_\mu^2 } {A_LA_R^*} {\Big \{ }
( I +   \gamma_5 \se)\kslash ~ 
 \gamma^0 \langle  \Eslash
 ~ P_R   ( I +   \gamma_5 \smu  ) \rangle \gamma^0  P_R 
~  \gamma^0 \langle \Eslash \rangle^*  {\Big \} }
\nonumber \\
&&-  {2G_F^2 m_\mu^2 } {A_L^*A_R} 
 {\Big \{ } ( I +  \gamma_5 \se ) \kslash ~
 \gamma^0   \langle \Eslash
  ~ P_L   ( I +  \gamma_5 \smu  )  \rangle \gamma^0   
~  P_L\gamma^0 \langle \Eslash \rangle^* {\Big \} }
\nonumber \\
&=& {8G_F^2 m_\mu^2 }
  {\rm Im} \{  A_LA_R^*\}
 |\langle \vec{E} \rangle|^2
\vec{s}_\mu \cdot (\vec{s}_e \times  \vec{k} ) 
\label{asym}
\eea
where  $\langle \vec{s}_\mu \cdot \vec{E} \rangle = 0$,
assuming a radial $\vec{E}$ field and all
the muons polarised.

To measure this process, suppose 
 that  the muon can be polarised by polarising the
target, as suggested in \cite{Kuno:1987dp}.
Consider  a coordinate system where
 $y-z$ is the plane of the
outgoing electron and the muon polarisation:
\beq
\hat{z} = \hat{s}_\mu ~~~~~~~~ \vec{k} = 
|\vec{k}| \sin \theta ~\hat{y}  +  
|\vec{k}| \cos \theta ~\hat{z} 
\eeq
Then one wishes to measure electron polarisation
in the perpendicular direction $\hat{x}$, which  means
in a cylinder aligned on
the polarisation direction of the muon.

For 100\% polarised muons, 
the asymmetry between electrons polarised in
the $+ \hat{x}$ and $- \hat{x}$  direction, 
normalised by the rate eqn (\ref{total}), is : 
\beq
\frac{2
\int d\Omega \vec{s}_\mu \cdot( \vec{s}_e \times \hat{k})}
{4 \pi}
\frac{{\rm Im} [A_L^* A_R]}{2 (|A_L|^2 + |A_R|^2)}
\leq \frac{{\rm Im} [A_L^* A_R]}{8 (|A_L|^2 + |A_R|^2)}
\label{what}
\eeq
which would be maximised for extensions 
of the SM that give similar, and complex,
contributions to $A_L$ and $A_R$.
The relative magnitude of $A_L$ and $A_R$
can be determined by the angular distribution
of the electron,  in $\mu \to e \gamma$, or  \mec~ 
providing the muon is polarised.

In summary, , Lepton Flavour Violating rates are an
important ingredient in determining
the New Physics at the TeV scale.  Experiments
are planned to improve the sensitivity of 
\mec ~  on nuclei, as compared to the rate
for nuclear $\beta$ decay by muon capture,  to
$10^{-16} - 10^{-18}$. If the muon could be
polarised (and the electron polarisation
is measured), a spin asymmetry in \mec ~
is sensitive to CP violating phases. This is shown
in this letter for the case where New Physics
induces  significant dipole interactions
(see eqn (\ref{eqn2})), but other dimension
six interactions are negligeable. We plan
a more complete analysis \cite{ACD?}.

\section*{Acknowledgements}

I thank Andrzej Czarnecki and
the organisers of $\nu$-Fact 08 for motivating
me to learn about muon physics, and 
A Czarnecki and N Rius for useful conversations.
This  work is supported by MEC-IN2P3 grant
IN2P3-08-05.

\end{document}